\documentclass[aps, prl, twocolumn, superscriptaddress, floatfix]{revtex4-2}
\usepackage{graphicx}
\usepackage{amsmath}
\usepackage{ulem}
\usepackage{amssymb}
\usepackage{tikz}
\usepackage{marvosym}
\usepackage{braket}
\usepackage{float}
\usepackage{pgfplots}
\usepackage{mathrsfs}
\usepackage{physics}
\usepackage[version=4]{mhchem}
\pgfplotsset{compat=1.17}
\usepackage{subcaption}
\usetikzlibrary{shapes.geometric}
\usetikzlibrary{shadows,patterns,perspective}
\usetikzlibrary {decorations,decorations.text}
\usetikzlibrary{decorations.pathreplacing}
\usepackage[pdfencoding=auto]{hyperref}
\usepackage{xcolor}
\definecolor{linkcolor}{RGB}{0, 0, 255}      
\definecolor{citecolor}{RGB}{0, 128, 0}     
\definecolor{urlcolor}{RGB}{255, 0, 0}       

\hypersetup{
    colorlinks=true,
    linkcolor=linkcolor,
    citecolor=citecolor,
    urlcolor=urlcolor,
    linktoc=all,  
    pdfborder={0 0 0}  
}

\DeclareCaptionJustification{justified}{\leftskip=0pt \rightskip=0pt \parfillskip=0pt plus 1fil}

\begin{document}
\definecolor{dy}{rgb}{0.9,0.9,0.4}
\definecolor{dr}{rgb}{0.95,0.65,0.55}
\definecolor{db}{rgb}{0.5,0.8,0.9}
\definecolor{dg}{rgb}{0.2,0.9,0.6}
\definecolor{BrickRed}{rgb}{0.8,0.3,0.3}
\definecolor{Navy}{rgb}{0.2,0.2,0.6}
\definecolor{DarkGreen}{rgb}{0.1,0.4,0.1}

\title{A mean-field description of strong-to-weak symmetry breaking in the monitored  
Bose-Hubbard model}

\author{Yicheng Tang}
\email{tang.yicheng@rutgers.edu}
\affiliation{Department of Physics and Astronomy, Center for Materials Theory,
Rutgers University, Piscataway, NJ 08854, United States of America}
\author{Pradip Kattel}
\affiliation{Department of Quantum Matter Physics, University of Geneva, Quai Ernest-Ansermet 24, 1211 Geneva, Switzerland}
\author{J. H. Pixley}
\affiliation{Department of Physics and Astronomy, Center for Materials Theory,
Rutgers University, Piscataway, NJ 08854, United States of America}
\affiliation{Center for Computational Quantum Physics, Flatiron Institute, 162 5th Avenue, New York, NY 10010}

\begin{abstract}
Strong-to-weak spontaneous symmetry breaking has emerged as a novel form of ordering in monitored and open quantum systems, yet its characterization has so far primarily relied on nonlocal diagnostics. Here, we develop a Gutzwiller mean-field framework for monitored bosonic lattice systems, enabling the direct simulation of stochastic measurement dynamics in three spatial dimensions. Applying this approach to the monitored Bose-Hubbard model with local density measurements and Lindbladian dissipation, we identify strong-to-weak symmetry breaking through a trajectory-averaged local order parameter. We find that this local order parameter becomes critical near the critical measurement strength as the charge-sharpening transition. Both transitions exhibit a dynamical exponent $z=2$ in the dilute limit with a correlation length $\nu\simeq 0.6$, comparable to that of the charge-sharpening transition, suggesting that the two phenomena may originate from a common underlying critical point. Our work shows a local characterization of strong-to-weak symmetry breaking, reveals its connection to charge sharpening, and provides concrete predictions for future experiments on the monitored Bose-Hubbard model.
\end{abstract}

\maketitle

\begin{figure}
    \centering
    \includegraphics[width=0.8\linewidth]{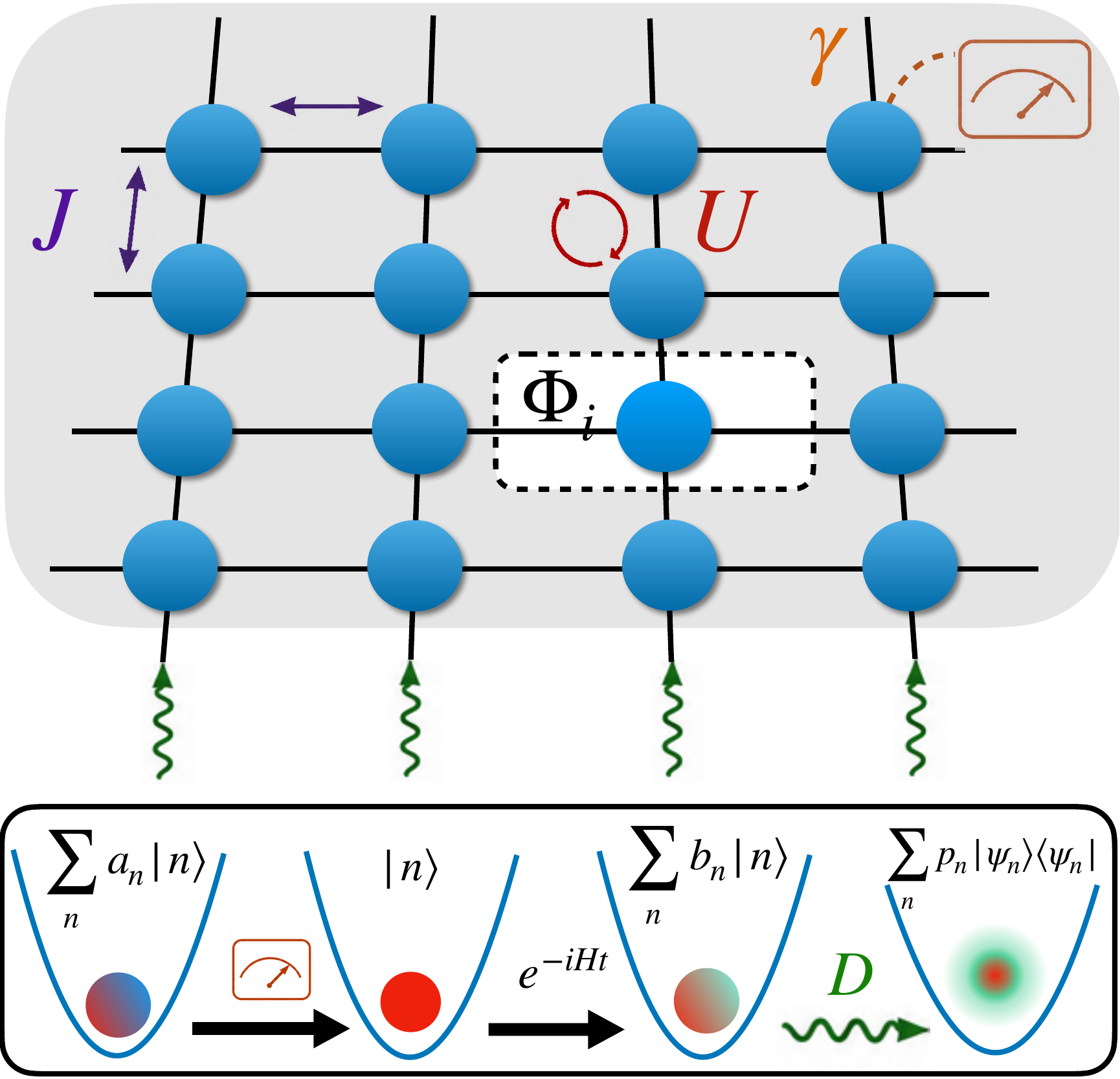}
    \caption{Illustration of the Bose-Hubbard model with hopping $J$ (blue arrows) and onsite interaction $U$ (red arrows), subject to measurement (orange dashed lines) at rate $\gamma$ and dissipation (green wavy lines) at rate $D$. The system is simulated using GMFT, where the local mean-field Hamiltonian at site $i$ (unshaded) is obtained by tracing out all other sites (shaded) and depends on the mean field $\Phi_i$ defined in Eq.~\eqref{eq:MFH}. The lower panel illustrates the interplay of measurement, Hamiltonian dynamics, and dissipation. Measurement projects the state onto a number eigenstate $|n\rangle$ (single-color circle), destroying quantum coherence. Subsequent Hamiltonian evolution regenerates coherence (multicolored circle). In contrast to measurement and coherent dynamics, which preserve purity, dissipation generates a classical mixture and drives the system into a mixed state described by a density matrix (fuzzy multicolored circle).}
    \label{fig:1}
\end{figure}

{\it Introduction}
Spontaneous symmetry breaking (SSB) is one of the central organizing principles of modern physics. It has successfully described the emergence of magnetism, crystallinity, superfluidity, and superconductivity in equilibrium, and provides a common language for describing phases that differ not in their microscopic constituents but in how they realize a  symmetry~\cite{landau1937theory,sachdev1999quantum}. In its conventional formulation, SSB concerns equilibrium systems or pure quantum states, where long-range order can be diagnosed through both correlation functions and local order parameters.
However, recently, mixed states have become unavoidable in quantum technologies. Modern quantum processors and memories are inherently open systems where noise, measurement, feedback, and error-correction protocols continuously generate and manipulate mixed states. Efficient descriptions of monitored quantum dynamics are essential for understanding device operation and for reaching the fault-tolerant regime. Tensor-network methods~\cite{schollwock2011density,bridgeman2017hand}, including matrix-product-state approaches, and stabilizer-based algorithms~\cite{gottesman1998heisenberg,aaronson2004improved,bravyi2016improved}, have provided powerful tools for qubit systems, but bosonic quantum-computing architectures remain considerably more challenging~\cite{shou2026anticoncentration,shou2026proof,deshpande2022quantum,oh2024classical,bouland2026complexity}. In particular, the long coherence times of photon modes in high-quality cavities make them promising candidates for oscillator-based quantum memories, while their infinite-dimensional onsite Hilbert spaces make the simulation of their monitored and dissipative mixed-state dynamics a major theoretical challenge~\cite{li2025cascaded}. Characterizing the symmetry structure of these mixed states could therefore be valuable across both quantum simulation and quantum information platforms~\cite{patel2026universal}.

This raises the question of how to reformulate the notion of symmetry breaking for open and monitored quantum systems. Recent developments have begun to address this question by extending the framework of symmetry breaking to nonequilibrium and mixed quantum states, where the symmetry structure of a density matrix contains information that is absent from the description of an individual pure state~\cite{buvca2012note,albert2014symmetries,sala2024spontaneous,lessa2025strong,gu2025spontaneous,huang2025hydrodynamics,4vs5-l54f,chen2025strong,hauser2026strong}. 
A particularly important distinction in mixed states is that between strong and weak symmetry. For a symmetry transformation with a unitary operator $U$ acting on a density matrix $\rho$, one can distinguish strong and weak symmetry. Strong symmetry, $U\rho=e^{i\theta}\rho$, constrains the ensemble to a definite symmetry-charge sector. Weak symmetry, $U\rho U^\dagger=\rho$, imposes invariance only under conjugation and permits different charge sectors to coexist~\cite{lessa2025strong}. A transition between these two structures is known as strong-to-weak spontaneous symmetry breaking (SWSSB). SWSSB is intrinsically a mixed-state form of ordering: it concerns not only the invariance of the density matrix, but also whether the ensemble itself has a sharply defined symmetry charge. Unlike conventional SSB, however, SWSSB has so far been characterized primarily through nonlocal information-theoretic quantities, including purification order parameters, fidelity-based diagnostics, conditional mutual information, charge-sector distinguishability, and learnability~\cite{sala2024spontaneous,lessa2025strong,4vs5-l54f,chen2025strong,singh2026mixed,hauser2026strong}. Whether this mixed-state order can be expressed through a local order parameter remains an important open question~\cite{zabalo2022nonlocality}.

Monitored quantum systems provide a natural setting in which to address this question. Measurements continuously acquire information about a quantum system, while coherent dynamics redistribute conserved quantities, generate entanglement, and compete with this information gain. Their competition can produce phases and critical phenomena with no equilibrium counterpart~\cite{li2018quantum,skinner2019measurement}. Ultracold atoms are particularly well suited to exploring this regime: optical lattices provide control over interactions and tunneling, while quantum gas microscopes enable site-resolved preparation and measurement of local occupations and correlations~\cite{jaksch1998cold,bloch2008many,gross2021quantum,cheneau2012light,huang2024two}. These capabilities have enabled studies of nonequilibrium dynamics, correlation spreading, and stochastic coherence, and have recently brought mixed-state symmetry breaking within experimental reach through the observation of SWSSB in a dephased Fermi gas~\cite{wang2026observation}. A central challenge is to identify local signatures of SWSSB that can be measured directly in a microscopic many-body experiment.

A closely related phenomenon arises in monitored systems with a conserved continuous charge. The competition between charge-conserving coherent dynamics and measurements can produce a charge-sharpening transition, separating a charge-fuzzy phase with substantial uncertainty in the conserved charge from a charge-sharp phase in which the charge sector becomes well resolved~\cite{agrawal2022entanglement,barratt2022transitions,barratt2022field,potter2022entanglement,majidy2023critical,oshima2023charge,agrawal2024observing,chakraborty2024charge,guo2025field,poboiko2025measurement,gopalakrishnan2026monitored}. From an information-theoretic perspective, charge sharpening is a learnability transition: in the charge-sharp phase, the measurement record contains sufficient information to identify the charge sector, whereas in the charge-fuzzy phase it does not~\cite{barratt2022transitions,singh2026mixed}. Since strong and weak symmetry differ precisely in whether distinct symmetry-charge sectors can coexist in the ensemble, charge sharpening and SWSSB appear to describe closely related aspects of the same measurement-driven reorganization. Establishing this connection would relate a previously nonlocal notion of mixed-state order to charge fluctuations and measurement records that are directly accessible in experiment.

Here we develop a Gutzwiller mean-field (GMFT) framework that makes this connection explicit for monitored interacting bosons and apply it to the three-dimensional Bose--Hubbard model with local density measurements and Lindbladian dissipation. GMFT provides more than a computational simplification in this setting. It reduces the monitored many-body problem to self-consistent stochastic single-site dynamics while retaining a local description of the quantities that diagnose charge sharpening and SWSSB. This makes it possible to investigate stochastic trajectories and critical scaling at system sizes relevant to optical-lattice experiments, while avoiding the direct evolution of a fully entangled many-body density matrix. Crucially, the approach turns SWSSB from an abstract classification based on nonlocal diagnostics into a phenomenon with a local order-parameter description and experimentally accessible signatures.

Within this framework, we identify a measurement-driven transition from a weak-symmetry, charge-fuzzy phase to a strong-symmetry, charge-sharp phase, both in the presence and absence of dissipation. The local SWSSB order parameter becomes critical at the charge-sharpening transition, and the two diagnostics exhibit consistent correlation-length scaling. These results provide evidence that charge sharpening and SWSSB are complementary manifestations of a common measurement-induced critical point rather than unrelated transitions. We further find diffusive critical dynamics, with dynamical exponent $z=2$ away from particle-hole symmetry, while a Lorentz symmetry with $z=1$ emerges at the particle-hole-symmetric point.

Our results establish a direct bridge between mixed-state symmetry breaking, charge learnability, and experimentally measurable local observables. They show that a symmetry structure previously diagnosed through nonlocal information-theoretic probes can be tied to charge fluctuations and local stochastic dynamics in a realistic interacting bosonic system. More broadly, GMFT provides a tractable route to studying mixed-state phases in systems where entanglement and infinite-dimensional onsite Hilbert spaces make exact many-body methods difficult. This combination of conceptual clarity, computational scalability, and experimental accessibility makes monitored bosons a promising setting for exploring mixed-state criticality and for developing a broader understanding of noisy quantum matter.

\begin{figure*}
    \centering
    \includegraphics[width=\linewidth]{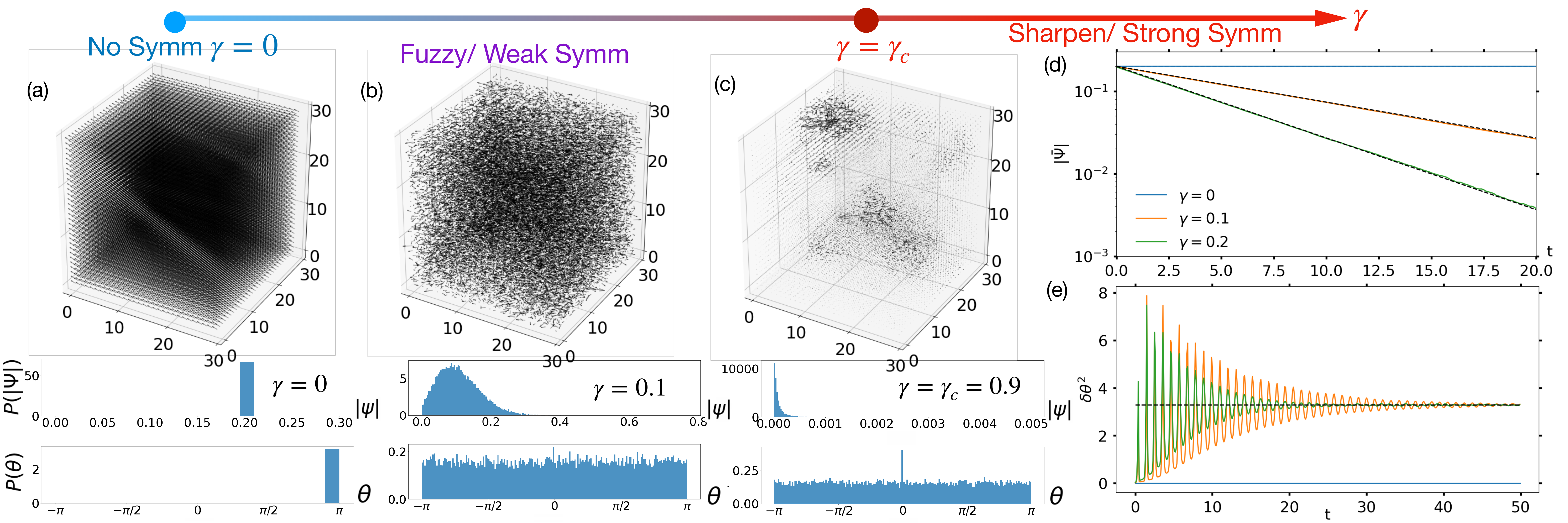}
    \caption{Panels (a)–(c) show the configuration of the superfluid order parameter $\Psi_i=|\Psi_i|e^{i\theta_i}$ with arrows for a single shot in the steady state $t=50$ with model parameter $D=0$, $U=J=1$, $\alpha=0.2$ and system size $L=30$ for various measurement rate $\gamma$. The norm $|\Psi|$ sets the length of a single arrow, and the arrow direction is determined by the phase in the direction $(\sin\theta,\cos\theta,0)$.  (a) $\gamma=0$ has broken symmetry with coherent phases $\theta_i=\theta_j$ and $\Psi_i=\Psi_j$ with a delta-function distribution in both variables over sites. (b) $\gamma=0.1$ in the weak symmetry (charge fuzzy) phase, where $|\Psi_i|\neq 0$ but with phases for each site that are randomly distributed. (c) $\gamma=0.9$ at criticality. To visualize the arrows, we enlarged the norm by a factor of $100$. Islands of weak symmetry at all scales occur at the criticality. Beyond the critical point $\gamma>\gamma_c$ is the charge sharp (strong symmetry) phase where $|\Psi_j|=0$ as a result of each site being in a number eigenstate. (d) shows for finite $\gamma$, although $|\Psi_i|\neq 0$, the averaged value $\bar{|\Psi|}(t)=|\sum_i\Psi_i|/V\simeq \alpha e^{-\gamma t}$ decays exponentially as a result of the phase dissipation, shown as the dashed lines. (e) shows the variance of the phase evolving to a uniform distribution with variance $\delta\theta^2=\frac{\pi^2}{3}$, indicated by a dashed line.}
    \label{fig:shot}
\end{figure*}

{\it Model and mean-field theory}
We consider the 3D Bose–Hubbard model subject to dissipation and continuous monitoring.
The dynamics consists of three components: the unitary part, the dissipation part, and the measurement part, as illustrated in Fig.~\ref{fig:1}.
The unitary dynamics is described by the Bose–Hubbard Hamiltonian defined on a cubic lattice of linear size $L$:
\begin{equation}
H=
-J\sum_{\langle i,j\rangle}
(b_i^\dagger b_j+\mathrm{h.c.})
+\frac{U}{2}\sum_j n_j(n_j-1),
\label{eq:bose-hubbard}
\end{equation}
where $\langle i,j\rangle$ denotes nearest neighbors, $b_j^\dagger$ ($b_j$) are bosonic creation (annihilation) operators satisfying $[b_i,b_j^\dagger]=\delta_{ij}$, and $n_j=b_j^\dagger b_j$ is the local number operator. At fixed filling, this Hamiltonian exhibits a quantum superfluid--Mott insulator transition as a function of $U/J$ at zero temperature~\cite{fisher1989boson,bloch2008many,greiner2002quantum,endres2012higgs}.

The dissipation is modeled by Hermitian Lindblad jump operators $L_i=n_i$ with the Liouvillian superoperator~\cite{lindblad1976generators,breuer2002theory}
\begin{equation}
    \begin{split}
        \mathcal{L}(\rho) = -i[H,\rho]+\sum_j\left[Dn_j \rho n_j-\frac{D}{2}\{n_j^2,\rho\}\right],
    \end{split}
\label{eq:Lindbladian}
\end{equation}
where $D$ is the dephasing rate, $\rho$ is the density matrix and $\{A,B\}=AB+BA$ is the anticommutator.

For the measurement part, the local density $n_j$ at each site is
continuously monitored at a rate $\gamma$. Such local density measurements are naturally motivated by quantum gas microscope experiments~\cite{bakr2009quantum,sherson2010single,gross2021quantum}. 
The monitored system evolves according to the stochastic Lindblad equation (SLE) as~\cite{daley2014quantum}
\begin{equation}
   \mathrm{d}\rho=
 \mathcal{L}(\rho) \mathrm{d}t+\sum_{n}
\sum_{j}
\left(
\frac{P_j^{(n)}\rho P_j^{(n)}}
{\langle P_j^{(n)} \rangle_\mathbf{m}}
-\rho
\right) \mathrm{d}N_j^{(n)},
\label{eq:SLE}
\end{equation}
where $\rho = \rho(t;\mathbf{m})$ is the density matrix at time $t$ subjected to a quantum trajectory $\mathbf{m} = \big(m(j_1,t_1),m(j_2,t_2)...\big)$ such that the site $j_i$ is measured at time $t_i$ with measurement outcome $m(j_i,t_i)$; $P_j^{(n)} = \ket{n_j}\bra{n_j} \otimes \mathbb{I}$ is the local number projector at site $j$; and the stochastic Poisson increments $\mathrm{d}N_j^{(n)}(t) \in \{0,1\}$ satisfy
$
\mathbb{E}\left[\mathrm{d}N_j^{(n)}(t) \right]
=
\gamma \langle P_j^{(n)} \rangle_\mathbf{m} \mathrm{d}t
$, where $\langle...\rangle_\mathbf{m}=\mathrm{Tr}\big(\rho(t,\mathbf{m})...\big)$ standing for quantum average with a given quantum trajectory $\mathbf{m}$ and$\mathbb{E}[...]$ denotes average over quantum trajectoies.  

Hilbert space dimension $\mathcal{H}_D = n_{\max}^{L^3}$, by truncating the local Hilbert space to dimension $n_{\max}$, becomes extremely large even for modest system sizes. 
To overcome this limitation, we employ a Gutzwiller mean-field theory (GMFT) approach~\cite{rokhsar1991gutzwiller,sheshadri1993superfluid}, which reduces the computational complexity to $\mathcal{O}(n_{\max}^2L^3)$, as detailed below. This computational efficiency enables direct access to system sizes and 3D geometries relevant to current optical-lattice experiments. The system is initialized in a local coherent state, $\rho(0)= \bigotimes_j |\alpha_j\rangle\langle \alpha_j|$ with $|\alpha_j\rangle = e^{-|\alpha|^2+\alpha b_j^\dagger}|0\rangle$. We introduce the MF Hamiltonian as 
\begin{equation}
    H_{\mathrm{MF}}[{\Phi}] = \sum_i \mathrm{Tr}_{\neq i}\Big(\bigotimes_{j\neq i}\rho_j(t,\mathbf{m})H\Big) = \sum_i h_i[\Phi_i],
\end{equation}
where  $\mathrm{Tr}_{\neq i}$ denotes tracing over all sites other than site $i$, and the MF is defined as $\Phi_i(t,\mathbf{m}) 
= \sum_{j\in \mathrm{nn}(i)} \mathrm{Tr}\Big(\rho_j(t;\mathbf{m})b_j\Big)$, with $\mathrm{nn}(i)$ being nearest neighbour of site $i$.
Thus, the local MF Hamiltonian reads
\begin{equation}
     h_j[\Phi_j] = - J \left( \Phi_j b_j^\dagger + \Phi_j^\ast b_j \right)+\frac{U}{2} n_j(n_j-1) .
     \label{eq:MFH}
\end{equation} 
Between two measurement events, the system evolves as
$\frac{d}{dt} \rho_j= 
-i[h_j,\rho_j]+Dn_j\rho_jn_j-\frac{D}{2}\{n_j^2,\rho_j\}$, where $\rho_j(t;\mathbf{m}) = \sum_{n,n'} \rho_j^{nn'}(t;\mathbf{m})|n_j\rangle\langle n'_j|$.
Here, the local MF Hamiltonian $h_j[\Phi_j]$ with $\Phi_j$ is updated at time $t$ as defined in ~\cite{PhysRevLett.119.073002}.
When site $j$ is measured with rate $\gamma$ according to the SLE defined in Eq.~\ref{eq:SLE}, the local density matrix is updated $ \rho_{j}^{nn'}(t,\mathbf{m}) \to \delta_{nm}\delta_{n'm}$, with Born probability $p_B(m) = \rho_j^{m,m}(t,\mathbf{m})$ subjected to measurement outcome $m$. Since the state remains unentangled $\rho(t;\mathbf{m}) = \otimes_j\rho_j(t;\mathbf{m})$ under GMFT, the many-body problem reduces to $L^3$ copies of a single-site boson simulation under the GMFT by neglecting intersite entanglement while retaining the full stochastic local dynamics generated by measurements and dissipation.

\begin{figure}
    \centering
    \includegraphics[width=\linewidth]{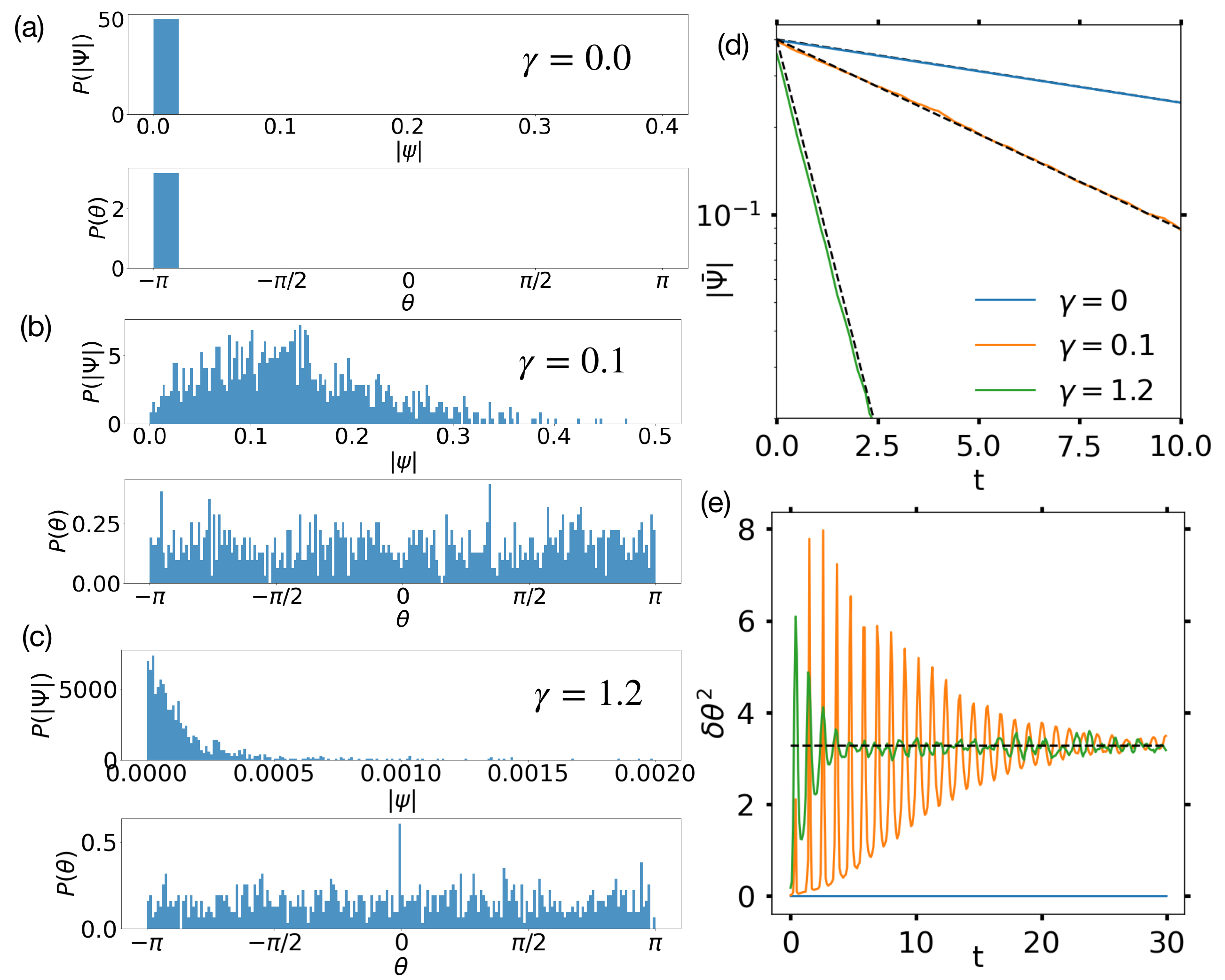}
    \caption{Configuration of the local superfluid order parameter $\Psi_i$ for a single quantum trajectory at $t=50$, with dissipation rate $D=0.1$, $U=J=1$, $\alpha=0.2$, and $L=30$. Panels (a)–(c) show the evolution from the weak-symmetry to the strong-symmetry regime as $\gamma$ increases. Unlike the nondissipative case, the system at $\gamma=0$ evolves toward the infinite-temperature mixed state $\rho_i\propto\mathbb{I}$, which is a weak-symmetry state. Panel (d) shows the decay of $|\bar{\Psi}(t)|\simeq \alpha e^{-(D/2+\gamma)t}$, indicated by the dashed line.}
    \label{fig:diss}
\end{figure}
\begin{figure*}
    \centering
    \includegraphics[width=\linewidth]{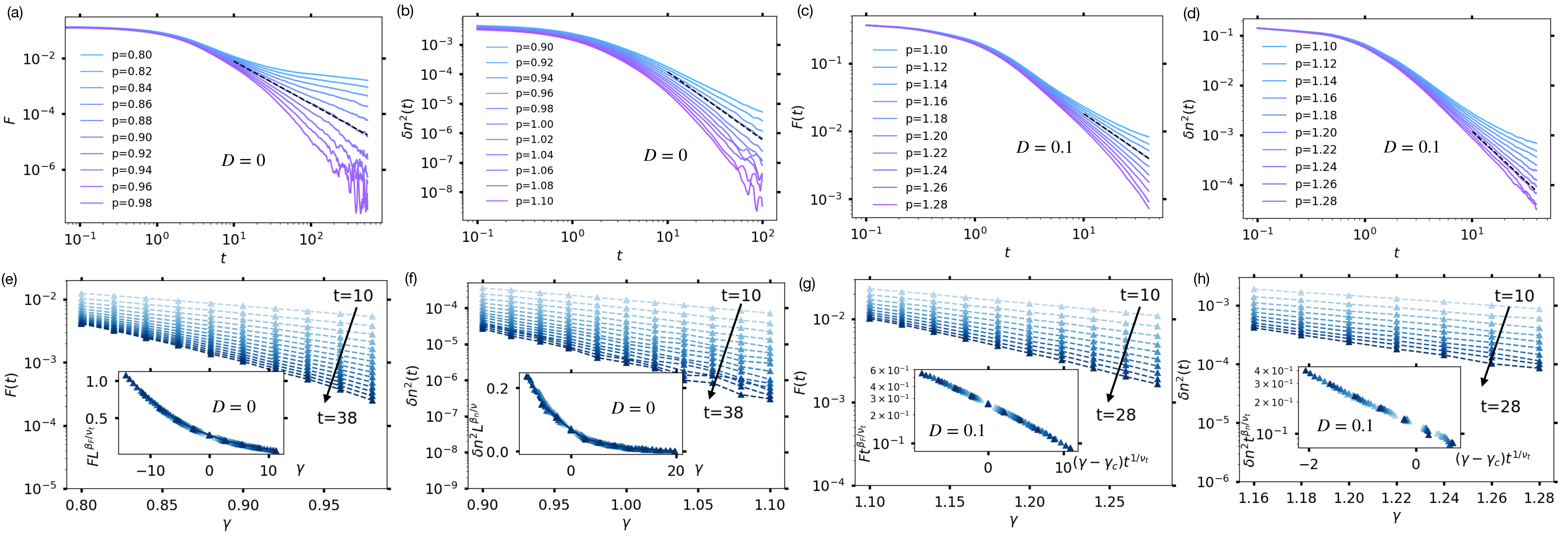}
    \caption{Upper four panels (a-d) show the time evolution for two order parameters, with varying measurement rate $\gamma$. (a) shows $F(t)$ with $\beta_F/\nu_t = 1.53\pm0.006$ and (b) shows $\delta n^2(t)$ with $\beta_n/\nu z = 2.31\pm 0.007$ at $L=30$, $\alpha=0.2$ and $D=0$.  (c) shows $F(t)$ with $\beta_F/\nu z = 1.10\pm0.005$ and (d) shows $\delta n^2(t)$ with $\beta_n/\nu z = 2.03\pm0.005$, at $L=16$, $\alpha=0.4$ and $D=0.1$. The dashed line is the fit of the scaling behavior at criticality $O(t) = \frac{A}{t^{\beta/\nu z}}$. The lower four panels (e-h) show the finite-time scaling behavior, with the insets showing the data collapse using the scaling ansatz defined in Eq.\ref{eq:scale}. (e) gives critical values $\gamma_c=0.89\pm0.01$, $\nu_t = 1.22\pm0.05$, (f) gives $\gamma_c=0.96\pm0.01$, $\nu_t = 1.20\pm0.05$ at $\alpha=0.2$ and $D=0$. (g) gives critical values $\gamma_c=1.18\pm0.01$, $\nu_t = 1.23\pm0.05$, (h) gives $\gamma_c=1.25\pm0.02$, $\nu_t = 1.32\pm0.04$ at $\alpha=0.4$ and $D=0.1$.
    All simulations here are done with parameters $U=J=1$, $n_{\max}=10$, and averaged over 1000 quantum trajectories.}
    \label{fig:scaling}
\end{figure*}

{\it Microscopic dynamics}: We begin by investigating the dynamics of a single trajectory $\mathbf{m}$. Without dissipation, $D=0$, the state remains pure and unentangled throughout the evolution $\mathrm{Tr}\big(\rho(t,\mathbf{m})^2\big)=\prod_i\mathrm{Tr}\big(\rho_i(t,\mathbf{m})^2\big)=1$.
To characterize the local coherence, we consider the superfluid order parameter, $\Psi_j = \mathrm{Tr}\big(b_j\rho_j(t,\mathbf{m})\big)=|\Psi_j|e^{i\theta_j}$, where the dependence on quantum trajectories $\mathbf{m}$ is left notationally implicit.  Without measurement, the system evolves coherently and spatially uniformly with $\Psi_j(t)=\Psi(t)$ as shown in Fig.~\ref{fig:shot}(a), and exhibits conventional symmetry breaking. Once the measurements are introduced, each measurement event projects the measured site to a number eigenstate, and destroys the coherence $\Psi_j\to 0$. Subsequent Hamiltonian dynamics will regenerate the coherence $\Psi_j =i\Phi_j \Delta t$ for a shot time interval $\Delta t$.  Therefore, when measurement events are rare, the measurement destroys the coherence more slowly than the generating process and results in the distribution of $|\Psi_j|$ peaking at a finite value with a non-vanishing mean in the thermodynamic limit (TDL). Nevertheless, the phases $\theta_j$ approach a uniform distribution on $[-\pi,\pi)$ as shown in Fig.~\ref{fig:shot}(b,c). As a consequence, the spatially averaged order parameter $\bar{\Psi}\equiv \sum_j \Psi_j/V=0$, where the overline denotes the site average. 
This behavior is reflected in the exponential decay $|\overline{\Psi}|(t)\simeq \alpha e^{-\gamma t}$ and in the phase variance approaching the value  $\delta\theta^2=\frac{\pi^2}{3}$ for a uniform distribution from $[-\pi,\pi)$, as shown in Fig.~\ref{fig:shot}(d,e). Thus, rare measurements destroy conventional symmetry breaking through phase decoherence rather than by eliminating local coherence. When the measurement occurs frequently, the local coherence is destroyed, resulting in a distribution of $|\Psi_j|$ with a vanishing mean in the TDL. 

When dissipation is turned on, $D\neq 0$, the qualitative effect of measurements remains unchanged. However, since the dissipation generates a classical mixture of pure states, the system is now in a mixed state which is described with a density matrix $\mathrm{Tr}\big(\rho_j(t,\mathbf{m})^2\big)<1$. When no measurement is applied, the dissipation drives the system to the infinite temperature steady state $\rho_i\propto\mathbb{I}$, which is a weak symmetry state with $\Psi_j=0$ as shown in Fig.~\ref{fig:diss}(a)~\cite{breuer2002theory}.  
For small measurement rates, the distribution of $\Psi_j$ acquires a non-zero mean value as shown in Fig.~\ref{fig:diss}(a-c), and vanishing mean with frequent measurement. The spatially averaged order parameter decays as $|\overline{\Psi}|\simeq \alpha e^{-(\gamma+\frac{D}{2}) t}$ while the phase is uniformly distributed as shown in of Fig.~\ref{fig:diss}(d,e).

{\it Phase Diagram}: 
The microscopic dynamics discussed above reveal that the measurements do not immediately eliminate local coherence. For weak monitoring, individual sites retain a finite local order parameter $\lim_{t\to\infty}\overline{|\Psi|}\neq 0$ in the steady state, whereas for sufficiently strong monitoring, local coherence is completely suppressed with $\lim_{t\to\infty}\overline{|\Psi|}= 0$ in the TDL.
This observation suggests the existence of two distinct phases: a weak-symmetry (charge fuzzy) phase with persistent local coherence and a strong-symmetry (charge sharp) phase where measurements fully localize the system.
Representative steady-state configurations of $\Psi_j$ at the steady state are shown in Fig.~\ref{fig:shot}(a-c) for $D=0$ and Fig.~\ref{fig:diss}(a-c) for $D=0.1$. At criticality, shown in Fig.~\ref{fig:shot}(c), islands of local coherence appear on all length scales, indicating critical fluctuations, which suggests that the loss of local coherence is associated with a genuine phase transition rather than a simple crossover

To characterize this transition quantitatively, we introduce the local Rényi-2 correlator
\begin{equation}
F(t,\mathbf{m}) = \overline{\sqrt{\mathrm{Tr}\Big(\sqrt{\rho (t,\mathbf{m})}b^\dagger_j \rho(t,\mathbf{m}) b_j\sqrt{\rho (t,\mathbf{m})}\Big)}},
\end{equation}
which is related to non-local descriptions of weak symmetry breaking such that the system is in a SWSSB phase if $\lim_{t\to\infty}F(t,\mathbf{m})\neq 0$\cite{lessa2025strong}. 
The physical meaning of $F$ is transparent without dissipation.
Since the state remains pure, the Rényi-2 correlator reduces to $ F(t,\mathbf{m})  =\overline{|\Psi_j(t)|}$. Therefore, $F$ directly measures the persistence of local coherence within a quantum trajectory, and when $|\Psi_j|\neq 0$, the system has a weak symmetry~\cite{buvca2012note}.
In contrast, when measurements are sufficiently strong ($\gamma>\gamma_c$), almost all sites are measured, leading to a strong symmetry phase. Unlike previously proposed nonlocal diagnostics of SWSSB, $F$ reduces to a local quantity within GMFT and suggests a route toward experimentally accessible probes based on site-resolved measurements.

The charge-sharpening transition is characterized by a different quantity, the charge variance
\begin{equation}
    \delta n^2(t,\mathbf{m}) = \frac{1}{V}\Big(\langle N^2\rangle_\mathbf{m}-\langle N\rangle_\mathbf{m}^2\Big),
\end{equation}
with total number operator $N = \sum_j n_j$. A finite value of $\lim_{t\to\infty}\delta n^2(t,\mathbf{m})\neq 0$ in the steady state indicates a charge fuzzy phase \cite{agrawal2022entanglement}. Within GMFT, the absence of intersite correlations implies that $\delta n^2(t,\mathbf{m}) = \overline{\delta n_j^2}(t,\mathbf{m})$ with local number variance $\delta n_j^2 = \langle n_j^2\rangle_\mathbf{m}-\langle n_j\rangle_\mathbf{m}^2$. 


\textit{Criticality:} 
We now test the central hypothesis that charge sharpening and SWSSB are two manifestations of the same underlying phase transition. To characterize their critical behavior, we compute the trajectory-averaged observables
\begin{equation}
    \delta n^2(t) = \mathbb{E}[\delta n^2(t,\mathbf{m})];\quad F(t) = \mathbb{E}[F(t,\mathbf{m})],
\end{equation}
which serve as the respective order parameters for charge sharpening and SWSSB transitions.
The simulation results are shown in Fig.~\ref{fig:scaling}.  We show that as time evolves, both parameters $\delta n^2(t)$ and $F(t)$ will saturate to a steady value, which is finite in the charge fuzzy (weak symmetry) phase and zero in the other phase in (a-d) of Fig.~\ref{fig:scaling} with and without dissipation. Near the critical point $\gamma_c$, all observables should obey the scaling ansatz, 
\begin{equation}
    O(t,L,\gamma) = L^{-\frac{\beta_O}{\nu}}f_O(L^{1/\nu}(\gamma-\gamma_c),t/L^z),
\end{equation}
or equivalently.
\begin{equation}
    O(t,L,\gamma) = t^{-\frac{\beta_O}{\nu_t}}f_O(t^{1/\nu_t}(\gamma-\gamma_c),t/L^z),
    \label{eq:scale}
\end{equation}
with $\nu_t=\nu z$ being the correlation time critical exponent.

At $\gamma=\gamma_c$, the relaxation of order parameters becomes algebraic, and both observables display critical power-law behavior, $ O(t) \propto \frac{1}{t^{\beta_O/\nu z}}$, with $O=F$ or $\delta n^2$. 
The data at criticality are well described by this scaling form, yielding $\beta_F/\nu z \simeq 1.353\pm0.006$ and $\beta_n/\nu \simeq 1.31\pm0.007$ for non-dissipative case and  $\beta_F/\nu \simeq 1.10\pm0.005$ and $\beta_n/\nu \simeq 2.03\pm0.007$ for dissipative case, and we find that the location of the critical point $\gamma_c$ between $F$ and $\delta n^2$ is reasonably close but different to each other as shown in Fig.\ref{fig:scaling} \cite{gu2025spontaneous}. Away from the particle-hole symmetric point, we find a dynamical exponent $z=2$, corresponding to diffusive critical dynamics. At the particle--hole symmetric point, however, the dynamical exponent becomes $z=1$, indicating emergent Lorentz invariance, as observed previously in other monitored systems~\cite{agrawal2022entanglement,barratt2022field}. The determination of the dynamical exponent and the distinction between the generic $z=2$ criticality and the particle--hole symmetric $z=1$ criticality are discussed in the Appendix.
To extrapolate the critical exponents, we apply the scaling ansatz
Eqn.\ref{eq:scale} with $t\ll L^2\to \infty $. The resulting data collapses are shown in (e-h) of Fig.~\ref{fig:scaling}, we find the correlation-length exponent of the SWSSB transition to be $\nu_t=1.22\pm0.05$ without dissipation and $\nu=1.23\pm0.05$ with dissipation, while for the charge-sharpening transition we obtain $\nu=1.20\pm0.05$ and $\nu=1.25\pm0.02$, respectively, as determined from the GMFT calculations.
The agreement between the correlation-length exponents extracted from $F$ and $\delta n^2$, both with and without dissipation, provides evidence that charge sharpening and SWSSB are governed by the same critical point. Our results further suggest that monitored Bose-Hubbard systems realized using quantum gas microscopes provide a promising route to experimentally observe charge sharpening and strong-to-weak symmetry breaking in interacting bosonic matter.

\textit{Conclusion:} 
We have developed a Gutzwiller mean-field framework for the monitored three-dimensional Bose--Hubbard model and identified a local description of strong-to-weak spontaneous symmetry breaking (SWSSB). Our results point to a broader connection between SWSSB and charge sharpening, which describe complementary aspects of monitored quantum matter: SWSSB characterizes how measurement changes the symmetry structure of the ensemble, whereas charge sharpening characterizes whether measurement resolves the conserved charge. The local Rényi-2 correlator becomes critical near the charge-sharpening transition, with the two transitions exhibiting consistent correlation-length scaling and critical dynamics characterized by $z=2$ away from particle-hole symmetry and $z=1$ at particle-hole symmetry.

These results provide evidence that SWSSB and charge sharpening may originate from a common underlying critical point rather than representing independent phenomena. They also show how mixed-state symmetry breaking, previously characterized primarily by nonlocal or information-theoretic diagnostics, can acquire a local manifestation via coherence and charge fluctuations, thereby demonstrating a direct connection among SWSSB, charge learnability, and experimentally accessible observables in site-resolved quantum-gas microscopy. GMFT omits intersite entanglement, leaving open whether this connection persists in fully entangled many-body states and extends to more general symmetries, potentially providing a broader framework for understanding measurement-induced phases and criticality.

Importantly, our framework provides an efficient route to describing correlations in the open-system dynamics of bosons, complementing recent work on monitored multimode photons in circuit QED~\cite{patel2026universal}. Developed here for monitored ultracold atoms, its stochastic single-site formulation offers a natural starting point for multimode cavity platforms, including recently demonstrated random-access quantum memories in circuit QED~\cite{li2025cascaded} and bosonic quantum-error-correction architectures~\cite{putterman2025hardware}.

{\it Note Added}: As this work was written up for publication, Ref.~\cite{lee2026charge,Divi:2026zrd,Zhang:2026nrp,Liu:2026lhj} appeared, where there is overlap, our results agree.

Acknowledgment: We thank Sarang Gopalakrishnan, David Huse, Udit Guntumadugu, Subhayan Sahu, and Romain Vasseur for insightful comments. This work is partially supported by the Army Research Office Grant No.~W911NF-23-1-0144  and the Rutgers Samuel
Marateck Fellowship. PK was supported by the Swiss National Science Foundation under
Division II (Grant No. 200020-219400). This work was partially conceived at the Kavli Institute of Theoretical Physics which is supported in part by the National Science Foundation under Grants No.~NSF PHY-1748958 and PHY-2309135 (J.H.P.)
\bibliography{ref}

\widetext

\newpage
\section*{Supplemental Material}
\subsection{Truncation of $n_{\mathrm{max}}$}
\begin{figure}[H]
    \centering
    \includegraphics[width=\linewidth]{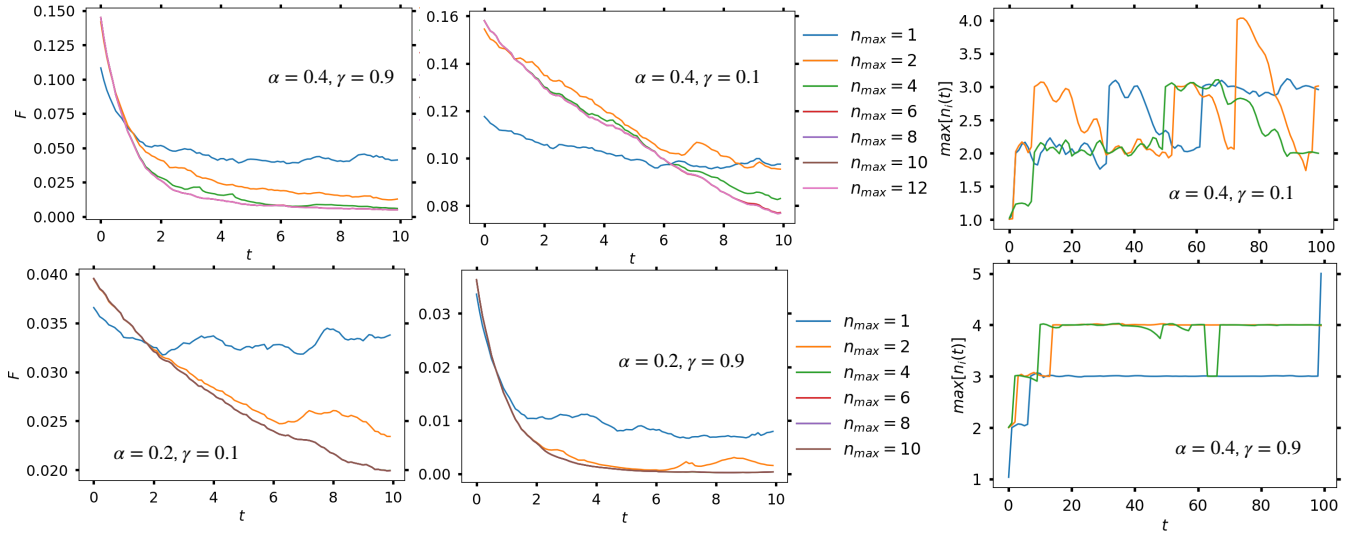}
    \caption{The left and middle panels show the time evolution of the Rényi-2 correlator $F(t)$ for different local Hilbert-space truncations $n_{\max}$, fillings $\alpha$, and measurement rates $\gamma$. The right panels show the corresponding maximum occupation number $\max[n(t)]$. All panels use the common color scheme indicated by the legend.}
    \label{fig:truncation}
\end{figure}

Here, we demonstrate the convergence of the dynamics with respect to the local Hilbert-space truncation $n_{\mathrm{max}}$. While small truncations $n_{\mathrm{max}}=1,2$ lead to visible deviations, the results rapidly converge with increasing $n_{\mathrm{max}}$. In particular, for the fillings $\alpha=0.2$ and $\alpha=0.4$ considered throughout the main text, the dynamics for $n_{\mathrm{max}}\geq 4$ are nearly indistinguishable on the scale of the figure. All simulations presented in the main text therefore use $n_{\mathrm{max}}=10$, well within the converged regime. Also, the maximum local occupation number for several quantum trajectories is shown in Fig.\ref{fig:truncation}. 

\subsection{Determination of the dynamical exponent}

\begin{figure}[H]
    \centering
    \includegraphics[width=\linewidth]{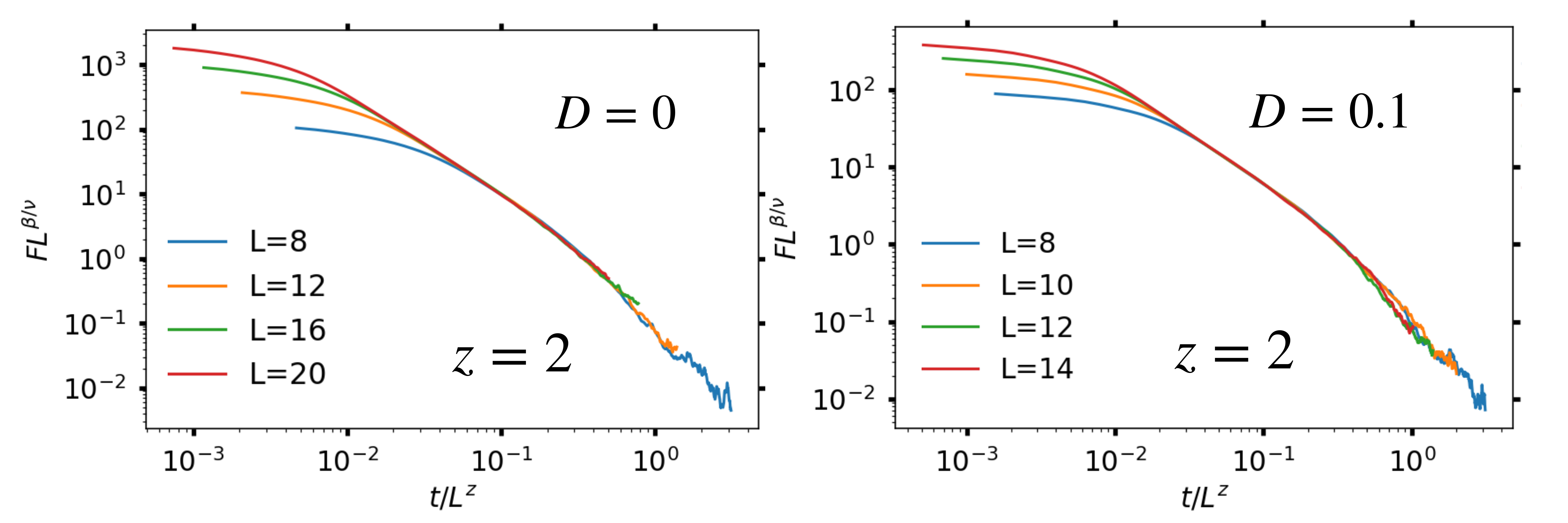}
    \caption{Data collapse of the dimensionless operators as a function of $t/L^z$ showing the dynamical exponent $z=2$.}
    \label{fig:zcollapse}
\end{figure}

At criticality, the order parameters obey the finite-size scaling form
\[
O(t,L)=L^{-\beta_O/\nu}f_O\left(t/L^z\right),
\]
where $O=F$ or $\delta n^2$, $\beta_O$ is the corresponding critical exponent, $\nu$ is the correlation-length exponent, and $z$ is the dynamical exponent. Therefore, the rescaled quantity $O(t,L)L^{\beta_O/\nu}$ becomes a universal function of $t/L^z$ at the critical point. As shown in Fig.~\ref{fig:zcollapse}, we obtain the best scaling collapse for $z=2$, consistent with diffusive dynamical scaling away from the particle--hole symmetric point.

\begin{figure}[H]
    \centering
    \includegraphics[width=\linewidth]{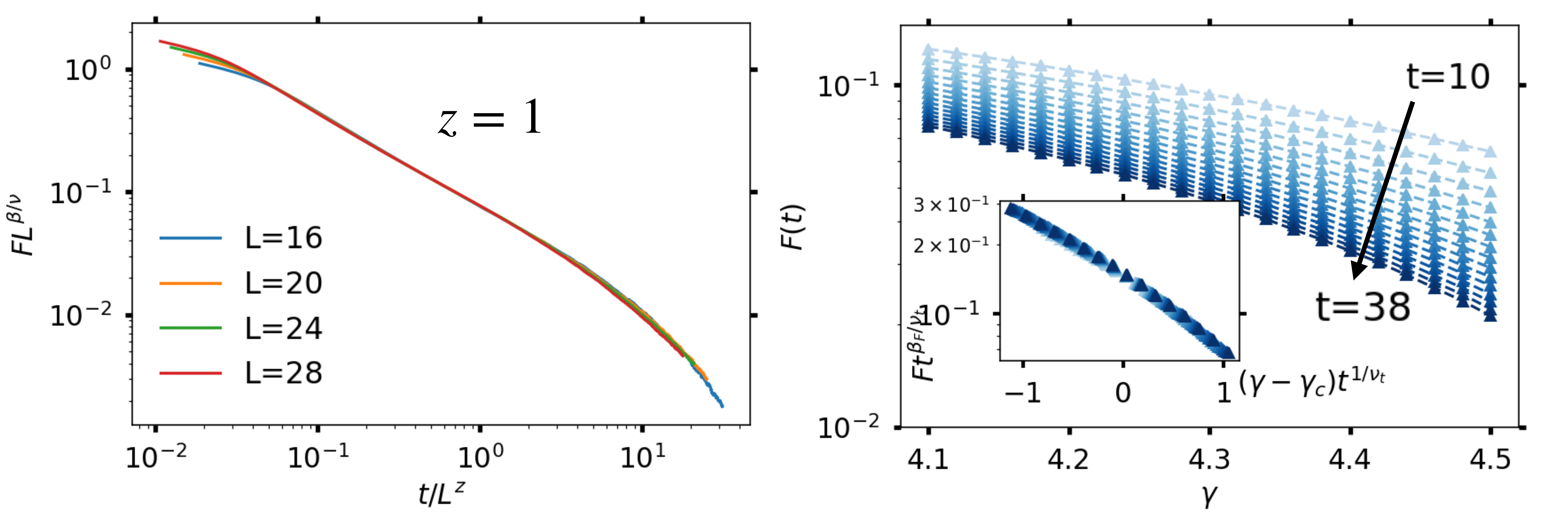}
    \caption{Left panel: Data collapse of the dimensionless operators as a function of $t/L^z$ showing the dynamical exponent $z=1$ at the particle-hole symmetric point. Right panel: finite time scaling for critical exponents $\gamma_c=4.25\pm0.01$ and $\nu_t = 2.6\pm 0.1$}
    \label{fig:zcollapse1}
\end{figure}

At the particle-hole symmetric point, we prepare the initial state as $|\psi\rangle = 0.1|0\rangle+|1\rangle + 0.1|2\rangle$ to make the state and the dynamics particle-hole symmetric after averaging over quantum trajectories. The dynamical exponent we extrapolated is $z=1$, corresponding to emergent Lorentz-invariant critical dynamics as shown in Fig.~\ref{fig:zcollapse1}. 

\end{document}